\documentclass[aps,preprint,showpacs,preprintnumbers,amssymb]{revtex4}

\usepackage[english]{babel}
\usepackage{amssymb}
\usepackage{indentfirst}
\usepackage{graphics}
\usepackage{fancyheadings}
\usepackage[T1]{fontenc}
\usepackage[dvips]{graphicx}
\usepackage{amsmath}
\usepackage{amsthm}

\usepackage{CatEn}

\usepackage{amsfonts,epsfig,color,bbm}

\begin{document}

\title{Algorithmic complexity of quantum states}

\author{C. Mora$^1$, H. J. Briegel$^{1,2}$}

\affiliation{$^1$ Institut f\"ur Quantenoptik und Quanteninformation
  der \"Osterreichische Akademie der Wissenschaften, Innsbruck,
  Austria\\
$^2$ Institut f{\"u}r Theoretische Physik, Universit{\"a}t Innsbruck,
Technikerstra{\ss}e 25, A-6020 Innsbruck, Austria} 

\date{\today}

\begin{abstract}

In this paper we give a definition for the Kolmogorov complexity of a
pure quantum state. In classical information theory the algorithmic
complexity of a string is a measure of the information needed by a
universal machine to reproduce the string itself. We define the
complexity of a quantum state by means of the classical description
complexity of an (abstract) experimental procedure that allows us to
prepare the state with a given fidelity. We argue that our definition
satisfies the intuitive idea of complexity as a measure of ``how
difficult'' it is to prepare a state. We apply this definition to give
an upper bound on the algorithmic complexity of a number of states. 
\end{abstract}
\maketitle

\section{Introduction and Notation}

Algorithmic information has provided a concise notion
of randomness for individual objects. It has also revealed deep
connections between thermodynamics and the theory
of classical computation \footnote{For a review see e. g. the paper by
  Bennett et al. \cite{tutti}}. The algorithmic complexity (or 
randomness) of an object -usually a binary string- is thereby 
defined as the length in bits of the shortest program for a universal
computer that reproduces the string under question \cite{kolmogorov}.
 
Quantum theory, on the other hand, has provided a new conceptual basis
for the theory of computation. Attempts have been made to
also generalize the notion of algorithmic complexity to quantum
mechanical objects, described by states in a Hilbert space. These
attempts are motivated primarily by the desire to formulate a
comprehensive theory of quantum information. We may also expect
further insights into the theory of entanglement.

In this paper, we want to characterize the algorithmic complexity of a
given quantum state. Proposals have already been made by Vitanyi
\cite{vitanyi} and 
Berthiaume et al. \cite{vandam} who have introduced two possible
definitions of 
quantum algorithmic complexity based on the reproducibility of the
state via Turing machines. G\`acs \cite{gacs} has instead adopted an approach
based on universal probability. Our definition will be closer to the
one proposed by Vitanyi, however deviating from his definition in two
crucial aspects. 

To introduce our definition of the algorithmic complexity of a quantum
state, we 
shall consider the following scenario. Imagine that Alice has created a
certain quantum state in her laboratory and wants to describe this
state to Bob, who wants to reproduce it in his laboratory. How
difficult is it 
for Alice to describe to Bob the state of her system? We may
distinguish the two situations in which Alice and Bob communicate via
a classical or a quantum channel, respectively. Depending on the
choice of the communication channel, we may arrive at different
notions of complexity of a quantum state. 

In the first situation, Alice has to use classical information to
describe her state to Bob. This appears to be a restriction, on 
first sight. However, we may always regard the quantum state
of a system as the result of some experimental preparation
procedure. The complexity of a quantum state is then associated, in a
very natural way, with the (classical) description complexity of an
experimental preparation procedure. The resulting notion of complexity
might therefore also be called \emph{preparation complexity} \footnote{This
 approach, might be closest to the traditional viewpoint -expressed
 most notably by Niels Bohr-  that the quantum state is essentially an
 expression of an experimental scenario.}. 

In the second situation, Alice may use quantum information to describe
her quantum state to Bob. In doing this, she has several options. She
may send either the quantum state altogether to Bob, or a copy (if
available), or the state in some Schumacher compressed form, or some
other quantum state which Bob can transform by into the desired
state. If we adopt this scenario, we arrive  -with Berthiaume
et al. \cite{vandam}- at a quite different notion of quantum complexity,
which might also be called \emph{encoding complexity}.

 In this paper, we shall follow the first approach and identify the
 algorithmic complexity of a quantum state with its preparation
 complexity i.e. the  classical description complexity of the
 preparation procedure. Although the second approach looks ``more
 quantum'', it lacks an important feature that we usually
 associate with the \emph{description} of an object. Even if Bob has the
 state sent by Alice in his hands, he might not know what state he has
 received. A proper description of the state, on the other hand, will
 allow him to reproduce the state himself

To be able to communicate, Alice and Bob must first have agreed on a
common language -- a non-trivial problem in linguistics -- which they
are using to describe their preparation procedure. Ideally, they will
use the same ``toolbox'' to compose their experiments and the
same words when referring to elements of this toolbox. In quantum
information theory, we abstract from a particular physical system in
which a quantum state is realized. The experimental toolbox is thereby
replaced by a set of elementary operations on a Hilbert space with a
given tensor-product structure and dimension \footnote{We would expect
  that any notion of complexity should be asymptotically invariant
  under coarsening of the description of the toolbox (for discussion
  of this point see Section \ref{coarsening}).}. The toolbox in quantum
information theory is thus a gross abstraction from an experimental
scenario. Here it the toolbox will include the possibility to prepare
some standard reference state, and a finite set of elementary
unitary transformations. A complete preparation procedure is then
described as a sequence of unitary transformations (a quantum circuit)
applied to the reference state.

Considering that a
quantum state can be characterized by a circuit with which the state
can be prepared, we want to define the complexity of a state referring
to that of the circuit itself. It is known that a finite state of gates
(constituting a complete basis) is suitable to prepare any state (up to
an arbitrary precision); through a sufficient coding, thus, the
circuit itself can be reduced to a (classical) string whose Kolmogorov
complexity is well defined and which can be associated to the original
state. In this way the algorithmic complexity of a state satisfies the
intuitive idea of complexity as a measure of ``how difficult'' it is
to prepare a state.

\vspace{0.2cm}

From now on we will represent with $\mcq_N$ the space generated by $N$
qubits and with $\mcq_N \ni |0\rangle = |0\rangle_N = |0\rangle_{(1)}
|0\rangle_{(2)}\cdots|0\rangle_({N}) $ the null vector (where
$|0\rangle_{(i)}\in\mcq_1^{(i)}$ is an element of the computational
basis $\{|0\rangle_1, |1\rangle_1\}$ of $\mcq_1^{(i)}$).

\noindent
We will represent with $\mcc|0\rangle$ the result of the application
of a circuit $\mcc$ on the null vector; if $\vert \sphi \mcc
|snul\vert^2 \geq 1 - \epsilon$ we will say that $\mcc$ prepares
$|\phi\rangle$ with precision $\epsilon$ (where $0\leq\epsilon \leq
1$). When saying that $\mcc$ prepares $\sphi$ we mean that
$\mcc\snul=\sphi$~. 

In particular we will be interested in building quantum circuits from
a fixed set of gates: as we want to be able to reproduce any state (at
least up to a given precision) this set must constitute a
\emph{complete gate basis}.

\begin{es}[Standard basis]
An example of a complete gate basis is the so called standard basis
\cite{nielsen} $\mcb = \{H,S,T,\cnot\}$, where $H$ is the Hadamard gate,
$S$ the phase gate, $T$ a $\frac{\pi}{8}$-gate and $\cnot$ the
controlled not: 
\best 
H = \frac{1}{\sqrt{2}}\left(\begin{array}{cc} 1
& 1\\ 1 & -1
    \end{array}\right)~;~~~~
S =\left(\begin{array}{cc} 1 & 0\\ 0 & i
  \end{array}\right)~;~~~~
T =\left(\begin{array}{cc} \esp^{-i\pi/8} & 0\\ 0 & \esp^{i\pi/8}
  \end{array}\right)~;~~~~
\begin{array}{cc}
    \cnot~\snul\sphi &=\snul\sphi\\
    \cnot~\sone\snul&=\sone\sone\\
    \cnot~\sone\sone&=\sone\snul
  \end{array}~.
\eest 
$T$ represents a $\pi/4$ rotation about the $z$ axis, while
$HTH$ a $\pi/4$ rotation around the $x$ axis.
\end{es}

Given a fixed (finite) number of gates only a
countable set of states can be prepared exactly. If we consider a
complete (finite) gate basis it is though possible to reproduce any
unitary transformation $\mcu$ (and thus any state $\sphi$) up to an
arbitrary precision. Considering that the definition of the
complexity of a quantum state will be based on its preparation
by means of a quantum circuit, it is therefore necessary to introduce
a \emph{precision parameter} in such definition. We will nevertheless
start defining the algorithmic complexity on the set of states that can be
exactly prepared with circuits built from a fixed basis. In this case
it obviously is not necessary to introduce this parameter; it will
appear only when generalizing this definition to arbitrary states.

\section{Classical Kolmogorov complexity}
\label{preliminary}

The definition of algorithmic complexity proposed by Kolmogorov
\cite{kolmogorov,chaitin1} is
meant to give an answer to the question: ``Is a (classical) sequence
random?''

Algorithmic complexity gives a definition of randomness very close to
the intuitive idea of ``structurless'' and is based on the concept of
\emph{algorithmic reproducibility} of a sequence. In practice, the
(classical) Kolmogorov complexity $K_\cl$ of a (binary) string $\bomega$ 
is defined as \emph{the length of the shortest program that, running
  on a universal Turing machine, gives $\bomega$ as output.} 

It follows quite easily from the definition that the algorithmic
complexity of a string $\bomega=\omega_{i_1}\omega_{i_2}\cdots$ can
grow at most linarily with the 
length of $\bomega$: it is in fact always possible to reproduce the
string by means of a program of the form: ``\emph{write $\omega_{i_1}$
  $\omega_{i_2}\cdots$}''. 

 This actually means that the length of a string constitutes
(up to a constant) an upper bound for the complexity of the
string itself:  
\beq
K_\cl(\bomega_n)\leq l(\bomega_n)+{\mathcal{O}}(1)=n+{\mathcal{O}}(1)
~. 
\label{complex}
\eeq

Naturally there are sequences for which this
upper bound is far too large: it is easilly shown, for example, that
the complexity of a periodic string grow only \emph{logarithmically}
with the length of the sequence.
A string is said to be \emph{complex} (or structureless, or
random) if its algorithmic complexity grows linearily with its
length: these are the strings typically generated by random sources
(such as, for example, a coin toss).

We will want to find a ``natural'' upper bound also for the complexity
of a quantum state. A
difficulty arises from the fact there seems to be no natural quantum
counterpart to the classical ``number of bits in the string''. We thus
find it necessary to look for another quantity, classically related to
the number of bits, for which such a counterpart exists. 

In order to do this, let us consider  the set of all infinite binary 
strings: it is easily shown that this set is isomorphic to the unit
interval $[0,1]$~\footnote{For each $\alpha\in[0,1]$ there exists in
  fact one (and only one) sequence $\{\alpha_i\}_i$ (with
  $\alpha_i\in\{0,1\}$) such that
$\alpha=\sum_i\alpha_i2^{-i}$.}. Through this isomorphism it is
possible to construct a (normalized) measure on the set of infinite
strings. Any $n$-bit binary string $\bomega_n$ identifies the set of infinite
strings whose first $n$ bits coincide with $\bomega_n$: the volume of
this set (a ball, $B_{\bomega_n}$) is $V(B_{\bomega_n})\sim 2^{-n}$.

The unit interval is thus divided in $[V(B_{\bomega_n})]^{-1}\sim 2^n$
subintervals, each identified by a $n$-bit sequence $\bomega_n^{(i)}$
with $i=1,2,\cdots,2^n$. Once we have numbered all the $n$-bit sequences
it follows immediately that each of them can be reproduced by a
program that specifies its index $i$, that is, by a program that
requires at most (up to some constant) $n=-\log 2^{-n} =
-\log V(B_{\bomega_n})$ bits. This simple
``counting'' argument gives an upper bound for the
complexity of an $n$-bit string which coincides with the one given by
equation (\ref{complex}):
\beq
K_\cl(\bomega_n)\leq-\log V(B_{\bomega_n})~.
\label{PrelComplV}
\eeq

The advantage of this reasonment is that it can be easily
adopted to find an upper bound to the complexity of quantum states. 

 In the quantum case, in fact, we will be looking at a circuit (or
 quantum 
 Turing machine or any other appropriate model) that reproduces a
 normalized quantum state $\sphi$ up to a fixed (given) precision
 $\epsilon$. This means that the circuit must prepare some quantum
 state $|\psi\rangle$ such that $\lvert
 \langle\psi\sphi\rvert^2\geq 1- \epsilon$: the set of all these states
 (any of which is acceptable as output for the circuit) defines a ball 
 in the $2^N$-dimensional space $\mcq_N$, with volume $V$ such that $V\sim 2^{-N}
 \epsilon^{2^N}$. This means that if $K^{\epsilon}(\sphi)$ is the
 complexity of the state $\sphi \in \mcq_N$ (when reproduced with
 precision $\epsilon$) we must have:
\beq
K(\sphi)\leq - \log V~\Leftrightarrow K^\epsilon(\sphi)\leq
-2^N\log\epsilon + N~. 
\label{PrelCondLin}
\eeq
In general the linear term can be omitted; we will thus usually
consider simply the condition:
\beq
K^\epsilon(\sphi)\leq -2^N\log\epsilon~.
\label{PrelCond}
\eeq

\begin{nota}
We underline that this is a preliminary condition, that should hold
true independently of the way one choses to define quantum algorithmic
complexity. It has in fact no relation to the model chosen to
reproduce the state, but depends instead only on \emph{a priori}
propertites such as the dimension of the space where the state is
defined and the precision with which the state must be reproduced.
\end{nota}

\section{Algorithmic complexity on a fixed set of states}
\label{coarsening}

In the following section we will assume to have fixed a complete gate
basis $B=\{G_1, G_2, \cdots, G_k\}$ and we will consider only states
$\sphi$ that can be prepared exactly by circuits built from $B$.

Once we fix a code (that is, an alphabet $\Omega = \{\omega_1, \omega_2,
\cdots, \omega_l\}$), the procedure to compute the algorithmic 
complexity of a state $\sphi$ the procedure is very simple.

\begin{enumerate}
\item With the gates contained in the basis $B$, build a circuit
  $\mcc^B(\sphi)$ such that $\mcc^B(\sphi)\snul=\sphi$~.
\item Code the circuit, obtaining a classical sequence 
  $\bomega^\Omega(\mcc^B) = \omega^\Omega_{i_1}
  \omega^\Omega_{i_2}\cdots\omega^\Omega_{i_m}$ of symbols
  $\omega^\Omega_{i_j}=\omega^\Omega_{i_j}(\mcc^B)\in\Omega$~. 
\end{enumerate}
\begin{nota}
Most of the code (that is, excluding some parts, e.g. a ``new line''
instruction or a way to identify the different quibits, that will be
more or less common to all codes) is strictly related to the gate
basis. In fact the code can 
be seen as a function that associates each gate of the basis to
a symbol (letter) or group of symbols (word). 
\end{nota}
\begin{enumerate}
\item[4.]{ We have now all the elements to define the algorithmic
  complexity of a state $\sphi$.
  \begin{defi} The algorithmic complexity of a state, relative to the  
  basis $B$, the code $\Omega$ and the circuit $\mcc^B(\sphi)$ is:
  \best
      K_\net^{\Omega,B,\mcc^B}(\sphi) =
      K_{\textrm{Cl}}(\bomega^\Omega(\mcc^B)) 
  \eest
  \label{NCcirc}
  \end{defi}}
\item[5.]{In general there are more circuits that prepare the same state
  $\sphi$, and in principle the correspondent complexities can
  be different. In order to define a property of the
  state itself (and not related  to the circuit used to reproduce it)
  we consider the following definition. 
  \begin{defi} 
    The algorithmic complexity of the state $\sphi$, relative to the code
    $\Omega$ and the gate basis $B$ is:
  \beq
       K_{\net}^{\Omega,B}(\sphi) = \min_{\mcc^B\in\tilde\mcc^B}
    K_{\net}^{\Omega,B,\mcc^B}(\sphi)
  \eeq
  where $\tilde\mcc_B$ is the set of all the circuits built with gates
  from $B$ that prepare $\sphi$.
  \end{defi}}
\end{enumerate}

Naturally, considering that the choices of code and basis are
arbitrary, it is necessary to study how they influence the complexity
of the state.

\vspace{0.2cm}

\begin{prop}[``Asymptotic'' invariance of the complexity of a state
  for code choice] 
If $\Omega$ and $\underline{\Omega}$ are two different codes, then, for
any state $\sphi$:
\beq
    K_{\net}^{\Omega,B}(\sphi) = K_{\net}^{\underline{\Omega},B}(\sphi) +
    k_{\Omega,\underline{\Omega}}~, 
\eeq 
where $k_{\Omega,\underline{\Omega}}$ is a constant that depends only on $\Omega$
and $\underline{\Omega}$.
\end{prop}
This can be seen as follows. 
For every $\mcc^B\in\tilde\mcc^B$, let
$\bomega^\Omega(\mcc^B)=\omega_{i_1}^\Omega \omega_{i_2}^\Omega \cdots 
\omega_{i_m}^\Omega$ and $\underline{\bomega}^{\underline{\Omega}}(\mcc^B) =
\underline{\omega}_{j_1}^{\underline{\Omega}}
\underline{\omega}_{j_2}^{\underline{\Omega}} \cdots 
\underline{\omega}_{j_n}^{\underline{\Omega}}$ be the strings that code
$\mcc^B$ using respectively codes $\Omega$ and
$\underline{\Omega}$.

$k_{\Omega,\underline{\Omega}}$ represents the length of a ``dictionary''
with which it is 
possible to translate the description made using one code to that made
using the other. Since both codes are finite, such dictionary is
finite too. The invariance is asymptotical since, in general,
$k_{\Omega,\underline{\Omega}}'$ can be very big and its relevance is 
lost only for $K_{\net}^{\Omega,B}(\sphi) \gg 1$~.

\begin{nota}
We underline that in general there is no corresponding invariance
property related to the basis choice (see
Sec. \ref{basis}). Nevertheless there are cases in which such an
invariance does hold true. This happens, for example, when we consider
a \emph{coarsening} of the gate basis, that is if we consider two gate
bases $B$ and $\underline{B}$, one of which
($\underline{B}$)constituted of non-elementary gates  that can be
built with gates from $B$ (e.g. $\underline{B}$ containes a Toffoli
gate, while $B$ contains Hadamard and C-not). In this case, in fact,
any circuit made by gates from $\underline{B}$ can be reproduced by
one made by gates from $B$. The string that codes this circuit will in 
general be longer than the one that of the original circuit, but their
complexities will change only for a (small) constant
$k_{B,\underline{B}}$ (that represent the length of a ``dictionary''
between the two gate bases).
\end{nota}

\vspace{0.2cm}

Considering the code-invariance property we can from now on omit
explicitating the dependence on the code (we can
imagine to fix it once and for all) and write simply:
$\displaystyle{K^B_\net(\sphi)}$~. 

\section{Algorithmic complexity for arbitrary states}

We want to generalize to arbitrary states what we have seen before. In
this case it is necessary to introduce the precision parameter
$\epsilon$: we can expect, in fact, that the greater the precision
with which the state must be reproduced by the circuit, the longer
will be the description of the circuit itself. 

\begin{nota}
The fact that the description of the circuit becomes longer does not 
necessarily mean that the complexity of the string that codes it(and
thus that of the state prepared by the circuit) 
increases. In fact, we can imagine some states that can be prepared
with better and better precisions by simply iterating the application
of a particular gate (or set of gates). In this case, the length of
the string that codes the circuit would naturally grow with the
precision, but not so the complexity of the circuit. However this
will not hold true in general, so it is necessary to keep the explicit 
dependence on the precision parameter. 
\end{nota}

The precision parameter enters in the definition of the algorithmic
complexity of the state $|\phi\rangle$ at the very first step, that is
in building the circuit that prepares it. When considering an
arbitrary state in the Hilbert space we will in fact need to specify
the precision up to which the circuit must prepare the state. We will
represent with $\mcc^B_\epsilon(\sphi)$ a circuit (built with gates
from $B$) that prepares $\sphi$ with precision $\epsilon$ and with
$\bomega^\Omega(\mcc^B_\epsilon(\sphi))$ the (classical) sequence that
codes $\mcc^B_\epsilon(\sphi)$.

\begin{defi}
The algorithmic complexity of state $\sphi$, relative to code $\Omega$ and
gate basis $B$ \emph{with precision parameter $\epsilon$} is:
\beq
   K_{\net}^{\Omega,B,\epsilon}(\sphi) =
   \min_{\mcc_\epsilon^B\in\tilde\mcc_\epsilon^B(\sphi)} 
      K_\cl(\bomega^\Omega(\mcc^B_\epsilon(\sphi))),
\eeq
  where $\tilde\mcc^B_\epsilon(\sphi)$ is the set of all the circuits
  built with gates 
  from $B$ that prepare $\sphi$ with precision $\epsilon$.
\end{defi}

\begin{nota}
The proof of the code-invariance of the complexity of a state, seen in 
the previous section, did not require that the state was reproduced
exactly by the circuits; thus the code-invariance
property still holds true. Therefore we can again we omit
explicitating the dependence on the code and write simply:
\beq
   K_{\net}^{B,\epsilon}(\sphi)~.
\eeq
\end{nota}

\section{Complexity and Precision}

In this paragraph we want to verify that our definition of a
complexity satisfies the preliminary condition given in Section 
\ref{preliminary}. In order to do this it is necessary to estimate the
upper bound of the algorithmic complexity of an arbitrary state $\sphi$.

It is known \cite{nielsen} that using only the (continuous) set of
all 1-qubit gates, plus the controlled not ($\cnot$), it is
possible to reproduce any unitary operation $U$ over $\mcq_N$ using
$\mco\left(N^2 4^N\right)$ gates. However, if one is interested to
reproduce the action of a unitary operation on one particular (given)
state only $\mco\left(N^2 2^N\right)$ such gates are sufficient; this
number of gates is thus sufficient to prepare any state $\sphi$ from
the given initial state $\snul$. 

We consider now the Solvay-Kitaev theorem\cite{kitaev} which implies
that any circuit acting on $\mcq_N$ built with $m$ 1-qubit gates and
$\cnot$'s  can be reproduced up to precision $\epsilon$ using 
$\mco\left(m \log^c\left(\frac{m}{\epsilon}\right)\right)$
gates from a finite gate basis ($c \in [1,2]$ is a constant whose
exact value is yet not known).

It follows immediatly that the action of any unitary transformation on
$\snul$ can be implemented (and thus any $\sphi\in\mcq_N$ can be
prepared) up to precision $\epsilon$ via a circuit built only with
gates from any finite and complete basis; futhermore, if $M$ is the
number of gates in the circuit, we have:
\beq
M=\mco\left(N^2 2^N \log^c \left(\frac{N^2
      2^N}{\epsilon}\right)\right)
      ~\Rightarrow~M\sim -N^2 2^N\log\epsilon~,
\label{numbergates}
\eeq
where the last expression is given considering only the leading order
in the two variables. 

Naturally the length of the string that codes the circuit grows
linearly with the number of gates of the circuit itself: in order to
code a circuit that prepares a general state $\sphi\in\mcq_N$ we need
thus a word whose length is (proportional to) $M$. Referring to the
definition given in the previous paragraph, in order to say that a
state $\sphi$ is complex it is necessary that its complexity (or,
equivalently, the complexity of the string
$\bomega^B_\epsilon(\sphi)$) grows linearily with the length ($M$)
of $\bomega^B_\epsilon(\sphi)$. From Equation (\ref{numbergates})
we obtain immediatly the logarithmic dependence on precision and the
exponential dependence on $N$ that were presented in Section
\ref{preliminary} as expected  upper bounds for the complexity: we 
have thus
\beq
K_\net^{B,\epsilon}(\sphi\in\mcq_N)\leq-N^2 2^N\log\epsilon~.
\label{UpperBound}
\eeq

\noindent
{\bf{Complex states}}

noindent
We are now in the position of defining what we mean by a
\emph{complex state}. Algorithmic complexity is an
uncomputable quantity: it is in fact based by definition on the
classical algorithmic complexity of the string $\bomega^\Omega(\mcc^B)$
that codes the circuit. As algorithmic complexity is an uncomputable
quantity \cite{chaitin1, kolmogorov, li} this property passes to algorithmic
complexity too.  

As defined above, the complexity of a state $\sphi$ is the minimum
complexity of a word that codes a circuit in
$\tilde\mcc^B_\epsilon(\sphi)$: this means that there is a circuit
$\hat{\mcc}^B_\epsilon(\sphi)\in\tilde\mcc^B_\epsilon(\sphi)$, coded
(using some alphabet) by $\hat\bomega^B_\epsilon(\sphi)$, such that 
$K_{\net}^{B,\epsilon}(\sphi) = K_\cl(\hat\bomega^B_\epsilon(\sphi))$.   

\begin{nota}
It is in general possible that there are more circuits that satisfy 
the same condition $K_{\net}^{B,\epsilon}(\sphi) =
K_{\net}(\hat\mcc^B_\epsilon(\sphi))$. In this case it is sufficient
to chose one of them: the choice is arbitrary and not relevant for the
following. 
\end{nota}

Once we have associated a classical string
(the \emph{characterizing string} $\hat\bomega^B_\epsilon(\sphi)$) to
the state $\sphi$ we can 
introduce the following definition:

\begin{defi}
A quantum state $\sphi\in\mcq_N$ is said to be \emph{complex} if the
classical string $\hat\bomega^B_\epsilon(\sphi)$ is complex.
\label{deficomplex}
\end{defi}

\noindent
As always, a classical string $\bomega$, of length $N$, is said to 
be complex if $K_\cl(\bomega)\sim N$. 

\begin{nota}
This definition satisfies the intuitive idea of the complexity of a
state. Let us in fact consider the following situation: Alice has
obtained a state ($\sphi$) and wants Bob to reproduce it (at least
with some precision $\epsilon$). Expecting that a
similar situation would arise, they had previously agreed on a
common code. All that Alice then has to pass to Bob is the information
on how to compose a circuit that prepares $\sphi$ with the given
precision, and this means sending Bob the string
$\bomega(\mcc^B_\epsilon(\sphi))$ that codes the circuit. In this case 
the complexity of the state $\sphi$ measures exactly the minimum
amount of information that Alice needs to pass to Bob. We underline
that this information \emph{is not} given by the length of the coding
string, but by its complexity: this simply reflects the fact that a
state could be prepared using a very big circuit (that will be coded
by a consequently long string), but a very simple one (again the
example of a circuit obtained repeating many times the same set of
gates): in this case the amount of information Alice needs will be
much smaller than the length of the coding word.
\end{nota}

\section{The ``basis problem''}
\label{basis}

The definition we have given for the algorithmic complexity of a state
has a very strong dependence on the choice of the basis. Fixing a
particular state it is in fact possible to build a particular basis so
that the description of $\sphi$ is trivial. One could thus argue that
the definition has no relevant meaning. 

Let us consider again the situation in which Alice prepares a state
and wants to describe it to 
Bob. If they have previously agreed on using a certain
gate basis, then Alice has only to describe to Bob the circuit (that
means passing to Bob the sequence $\bomega_\mcc$). If they have not
agreed on a particular gate basis, then Alice could indeed build a
circuit using the ``best'' basis, but in this case she would have to
describe the basis itself to Bob, and this would be in general a
similarly difficult task \footnote{This, in fact, would again require
  to describe arbitrary unitary transformations.}. 

One might nevertheless wonder whether there is some particular basis
that allows to describe all (or almost all) states with simple
circuits. If such a basis existed, it would obviously be convenient
for Alice and Bob to agree on using that: in this case we would obtain
that (almost) all quantum states are non-complex. In the following we
will show that such a basis cannot exist as, once \emph{any} gate
basis is fixed , the number of non-complex states is always small in
relation to the total number of states. 

In classical information theory it is well known that the number of
compressible (bit) strings is ``small''; more precisely one has: 
\beq
\frac{\#\{\bomega_n=\omega_{i_1}\cdots\omega_{i_n} |
  K_\cl(\bomega_n\} < c\}}
{\#\{\bomega_n=\omega_{i_1}\cdots\omega_{i_n}\}} \leq
\frac{2^c-1}{2^n}~.  
\label{cardinalitycl}
\eeq

Now let us consider the quantum case: as we have seen in the previous
paragraphs, once we fix a basis $B$ and a
precision parameter $\epsilon$, we can associate to every quantum
state $\sphi$ a $(-N^2 2^N\log\epsilon)$-bit string
$\hat\bomega_\epsilon^B(\sphi)$ whose classical algorithmic complexity
coincides with the complexity  of $\sphi$. Applying
Eq. (\ref{cardinalitycl}) to the set of strings $\hat\bomega_\epsilon^B$
we obtain:
\beq
\frac{\#\{\sphi\in\mcq_N |
  K_\epsilon^B(\sphi)<c\}}{\#\{\sphi\Big\vert_\epsilon\}}\leq 
\frac{2^c -1}{2^{-N^2 2^N\log\epsilon}}\simeq 2^{N^2 2^N \log\epsilon+c}~,
\label{cardinalityq}
\eeq
where with $\#\{\sphi\Big\vert_\epsilon\}$ we represent the number of
different normalized states that can be prepared with precision
$\epsilon$. Such a relation holds true also in the case when
$c=c(N,\epsilon)$ is a function.
State $\sphi$ will be non-complex only if its complexity is
$o(-N^2 2^N\log\epsilon)$: this means that the right member of the
inequality
becomes:$2^{N^2 2^N\log\epsilon+o(-N^2 2^N\log\epsilon)}\sim 2^{N^2
  2^N\log\epsilon}\ll 
1$. Thus, applying equation (\ref{cardinalityq}) to
these states, one obtains that for any fixed basis $B$, the number of
non complex states is exponentially small. 

\section{Entropy and complexity}

In classical information theory one can consider a random source
that emits (with a certain probability distribution) letters drawn
from some finite alphabet. In this case there is a strong relationship
between the Shannon entropy of the source and the 
algorithmic complexity of the emitted sequences. In particular, if
$\bomega_n$ is a $n-$letter sequence and $p(\bomega_n)$ is its
probability, one obtains \cite{gruenwald}:
\beq
\vert \langle K_\cl(\bomega_n) - H \rangle \vert \leq c~,
\label{complexentro}
\eeq
where the average is taken over all $n$-letter sequences, $H =
-\sum_{\bomega_n} p(\bomega_n)\log p(\bomega_n)$ is the Shannon
entropy of the source and $c$ is a constant that depends on the
probability distribution. When such a distribution does not depend on
the length of the sequences (as in the case of Bernoulli sources),
this implies that, in the limit of $n \gg 1$, the average complexity
production of the source coincides with its entropy production:
\beq
\frac{\langle K_\cl(\bomega_n)\rangle}{n} \xrightarrow{n\to\infty}
\frac{H}{n}~. 
\label{complexentrorate}
\eeq

It comes quite natural to seek a similar relation in the context of
quantum information theory. In order to do this, we will first of all
find the corresponding relation in the case of a classical source that
emits words (and not letters). After that we will define what we
intend by quantum source and find, in this case, the wanted
relationship between complexity and entropy. 

\subsection{Classical case}
\label{sentences}

Let us consider now a variation of the classical letter-source: in this
case we will have a source that emits, with some given probability,
\emph{words} drawn from a finite dictionary $D$. Each of the words will be
a sequence of letters of alphabet $A$; without loss of generality we
can assume all words to have the same length $l$. The output of such a
source will thus be a sequence of words (or
\emph{sentence}). As the Shannon entropy depends only on
the probability distribution, once we fix such a distribution it is
the same for letter and for word-sources. If we consider the
complexity of the emitted message, though, it is evident that there
must be some differences. A word-source that emits $m$ objects will in
fact have transmitted a sequence of $m l$ letters: it is now easy
to believe that the complexity of such a sequence can be higher than
that of an $m$-letter sequence (even though letters and words are
emitted following the same probability distribution). This is an
immediate consequence of the fact that words are composite objects
that have non-zero complexity themselves. 

Naturally, when we consider very long output sentences, such a
difference becomes negligible. Any sentence can in fact be reduced to
a word by a program that associates to each word a symbol: the length
of this program will be determined by the complexity of the dictionary
(that is, by the complexity of the single
words in the dictionary) and will thus be bound by the dictionary
length $l \#D$. This contribution, though, can be extremely relevant
while we consider ``short'' sentences, that is sentences whose length
is comparable to that of the dictionary. 

Thus, if $\boldsymbol{\Omega}_m=\bomega_{i_1}\cdots\bomega_{i_m}$ is
an $m-$word sentence, its complexity is given by:
\beq
K_\cl(\boldsymbol{\Omega}_m) \simeq
K_\cl(\boldsymbol{i}_m= i_1\cdots i_m) +
\sum_{j=1}^{\#D}K_\cl(\bomega_j) \leq
K_\cl(\boldsymbol{i}_m) + l \#D~,
\label{sentcompl}
\eeq
where $i_k$ is the symbol that codes $\bomega_{i_k}$, and $l$
is the common length of the words $\bomega_k\in D$~.

The sequences $\boldsymbol{i}_m$ can easily be seen as the 
outputs of a letter-source whose entropy $H$ coincides with that of
the considered word-source: for such sequences, thus, the relation
expressed by Eq. (\ref{complexentro}) still holds true. When
generalizing it to word-sources, though, it is necessary to consider
the contribute of the dictionary, obtaining:
\beq
\vert \langle K_\cl(\boldsymbol{\Omega}_m )- H\rangle -
\sum_{j=1}^{\#D}K_\cl(\bomega_j)\vert \leq c_1~.
\label{sentencecomplexity}
\eeq
Naturally, being the complexity of the dictionary constant,
Eq. (\ref{complexentrorate}) remains of the same form for word-sources
too:
\beq
\frac{\langle K_\cl(\boldsymbol{\Omega}_m)\rangle}{m}
\xrightarrow{m\to\infty} \frac{H}{m}~.
\eeq

\subsection{Quantum case}

Before being able to say anything for the quantum case, it is
necessary to specify what we will consider as a quantum source. In
principle, any mixed state can be viewed as a quantum source. In the
following, though, we will consider a quantum source as a 
``black box'' that emits (with a given probability) pure states drawn
from a given set \cite{schumacher}. This corresponds to considering
not a general mixed 
state, but rather a well defined mixture
$\{(p_j,|\phi_j\rangle)\}_{j\in D}$ of pure states ($|\phi_j\rangle
\in \mcq_N$). In this case
the source has a kind of ``semi-classical'' nature: it is in fact
quantum only in the sense that it emits quantum states and not (as in
the cases conidered in the previous paragraphs) classical objects,
while it cannot, for example, emit states that are other than the
tensor product of the ones in the ensemble. For these sources the
Shannon entropy coincides with the Shannon entropy of a 
classical source that emits objects (letters or words) with the same
probability distribution $\{p_j\}_{j\in D}$. 

As we have seen when defining the complexity of a quantum state, once
we fix a finite precision $\epsilon$, it is possible to describe any
quantum state by means of a finite word (whose length depend both on
the dimension of the Hilbert space in which the state lives and on the
value of $\epsilon$). This fact allows us to identify a quantum source
of this kind with a corresponding word-source. The expression for the
complexity of the emitted messages (tensor-product states) follows
thus immediately from what seen above: 
\beq
\begin{split}
K_\net^{B,\epsilon}(|\boldsymbol{\Phi}_m \rangle & =|\phi_{i_1}\rangle 
|\phi_{i_2}\rangle \cdots |\phi_{i_m}\rangle) \simeq
K_\cl(\boldsymbol{i}_m= i_1\cdots i_m) + 
\sum_{j\in D}K^{B,\epsilon}_\net(|\phi_j\rangle) \\
& \leq K_\cl(\boldsymbol{i}_m) - \#D N^2 2^N\log\epsilon~, 
\end{split}
\label{stateword}
\eeq
where $i_k$ is the symbol that codes state $|\phi_{i_k}\rangle$.
We underline that, in this case, the contribution due to the
complexity of the dictionary can be relevant indeed, being bound by
the complexity of the states (that can in principle be very large). 
 The relation between the entropy $H$ of the source 
(that is, the Shannon entropy of the probability distribution
$\{p_j\}_j$) and the complexity of the message is analogously obtained: 
\beq
\vert \langle K_\net^{B,\epsilon}(|\boldsymbol{\Phi}_m\rangle) -
H\rangle - \sum_{j\in D}K^{B,\epsilon}_\net(|\phi_j\rangle) \vert \leq
c~.  
\eeq

As in the classical case, though, if we consider the limit of
infinitely long state sequences the complexity rate and the entropy
rate tend to coincides: in this limit, thus, the source ``reveals''
its semiclassical nature.

\begin{nota}
Quite naturally one coul ask what happens if we have the possibility
to apply Schumacher's noiseless coding theorem \cite{schumacher} and
thus compress the emitted states. If $S(\rho)$ is the von Neumann
entropy of the source $\rho=\sum_jp_j|\phi_j\rangle\langle\phi_j|$,
each state can be compressed into a new state $|\phi_j'\rangle$ in a
$(2^{N S(\rho)})$-dimensional Hilbert space. In this case we can
rewrite Eq. (\ref{stateword}) as:
\beq
\begin{split}
K_\net^{B,\epsilon}(|\boldsymbol{\Phi}_m \rangle & =|\phi_{i_1}\rangle 
|\phi_{i_2}\rangle \cdots |\phi_{i_m}\rangle) \simeq
K_\net^{B,\epsilon}(|\boldsymbol{\Phi'}_m \rangle =|\phi_{i_1}'\rangle  
|\phi_{i_2}'\rangle \cdots |\phi_{i_m}'\rangle) \\ 
& \simeq K_\cl(\boldsymbol{i}_m= i_1\cdots i_m) + 
\sum_{j\in D}K^{B,\epsilon}_\net(|\phi_j'\rangle) \\
& \leq K_\cl(\boldsymbol{i}_m) - \#D (NS(\rho))^2 2^{N
  S(\rho)}\log\epsilon~. 
\end{split}
\eeq
\end{nota}

\section{Applications and examples}

\subsection{Complexity of copies}

One of the properties of classical algorithmic complexity is that
obtaining $m$ copies of a given string is (almost) free, being the
dependence of the complexity on the number of copies only logarithmic:
\best
K_\cl(\bomega^{(m)}=\underbrace{\bomega
  \bomega\cdots\bomega}_{m~\textit{times}}) 
\leq K_\cl(\bomega)+ {\mathcal{O}}(\log m)~.
\eest
This bound is easilly obtained by considering that, once one has a
program that reproduces string $\bomega$ it is sufficient to run it
$m$ times to reproduce $\bomega^{(m)}$.

In the case of a quantum system, the situation is slightly more
complex. It is first of all necessary to decide exactly what we mean
by ``preparing $m$ copies'' of a quantum state: in fact
either we will require the circuit to prepare $m$ times a quantum
state $\sphi\in\mcq_N$, with some fixed precision parameter $\epsilon$,
or it must prepare the global state $\sphi^{\otimes m} \in
\mcq_N^{\otimes m}$ with the
given precision. The two situations are extremely different. 

In the first case, we obtain the same relation we have seen in the
classical case:
\best
K_\net^B(\sphi\Big\vert_\epsilon^{(m)}) =
K_\net^B(\underbrace{\sphi\Big\vert_\epsilon 
  \cdots \sphi\Big\vert_\epsilon}_{m~\textit{times}}) \leq
K_\net^{B,\epsilon}(\sphi)+\mco(\log m)~,
\eest
where the expression $\sphi\Big\vert^{(m)}_\epsilon$ wants to remind
us that each copy 
of $\sphi$ is reproduced with precision $\epsilon$~. This is an
immediate consequence of the definition of complexity: a circuit that
prepares $\sphi\big\vert_\epsilon^{(m)}$ can be in fact built by
repeating $m$ 
times the one that prepares $\sphi$ with precision $\epsilon$.

If, instead, we require the state $\sphi^{\otimes m}$ to be prepared
with precision $\epsilon$, the situation is different and the
classical relation does not (necessarily) hold true any more. In this
case the relation between the complexity of $\sphi$ and that of
$\sphi^{\otimes m}$ has the following form: 

\beq
K_\net^{B,\epsilon}(\sphi^{\otimes m}) \leq
K_\net^{B,\frac{\epsilon}{m}}(\sphi) + \mco (\log m)\lesssim -N^22^N
\log \frac{\epsilon}{m}~.
\eeq

This expression follows immediately by the fact that the state
$\sphi^{\otimes m}$ can be prepared with precision $\epsilon$ by a
circuit $\mcc^B_\epsilon(\sphi^{\otimes m)}$ built with $m$ identical
copies of a smaller circuit $\mcc^B_{\epsilon/4m}(\sphi)$ \cite{holevo}.\\
The last inequality follows immediately from
Eq. (\ref{UpperBound}). We underline that in most cases  (that 
is when $4 m \epsilon^{m-1}\leq 1$) this bound is much stricter
than the one we could obtain by directly applying (\ref{UpperBound})
with which one has: $K_\net^{B,\epsilon}(\sphi^{\otimes m}) \leq -m
N^2 2^N\log\epsilon$~. 

\subsection{Entanglement and Complexity}

Let us consider a state $\sphi \in \mcq_N$; suppose it can be
written as
\best
\sphi = |\phi_1\rangle \otimes |\phi_2\rangle\otimes \cdots
|\phi_J\rangle,~~\textrm{with}~|\phi_j\rangle\in\mcq_{N_j}~\textrm{such
that}~\textit{dim}(\mcq_{N_j})=2^{N_j},~\textrm{and}~\sum_{j=1}^J
N_j=N~. 
\eest
This means that the state $\sphi$ is not totally entangled, and it can
thus be considered as the tensor product of other (possibly entangled)
states $|\phi_j\rangle$ .

As a consequence of this fact we have:
\beq
K_\net^{B,\epsilon}(\sphi) \leq
\sum_{j=1}^J K_\net^{\epsilon/J}(|\phi_j\rangle) \leq -\sum_{j=1}^J
N_j^2 2^{N_j}\log\frac{\epsilon}{J}~.
\eeq
We want to show that this upper bound is actually stricter than the
one obtained for general states in $\mcq_N$: let us rewrite the bound
given by Eq. (\ref{UpperBound}) as
$\displaystyle{K_\net^{B,\epsilon}(\sphi) \leq
-\sum_{j=1}^J \frac{N^2 2^N}{J}\log\epsilon}$. The wanted inequality is proved
by considering:
\best
N_j^2 2^{N_j}\log\frac{J}{\epsilon} < N^2
2^{N_j}\log\frac{J}{\epsilon} \leq 
\frac{N^2 2^N}{2^{J-1}}\log\frac{J}{\epsilon} <
\frac{N^2 2^N}{J}\log\frac{1}{\epsilon}~.
\eest

Thus, the \emph{maximal complexity} can be obtained \emph{only} by a
truly $N$-party entangled state (in the sense that it cannot be written as
tensor product of states contained in subspaces of $\mcq_N$). We
stress that this consideration does not imply that all totally
entangled states have maximal complexity (as a counterexample it is
enough to consider the GHZ states).

It is nevertheless interesting to consider how a property that is
characteristic of quantum systems (that is entanglement) has direct
effect on the complexity of a state.

\begin{es}[Complexity of a completely separable state]
As an example let us consider a state $\sphi\in
\mcq_N$ of the form $\sphi = \bigotimes_{j=1}^N
|\phi_i\rangle,~~|\phi_i\rangle\in\mcq_1$. In this case we have:
\best
K_\net^{B,\epsilon}(\sphi)\leq\sum_{j=1}^N
K_\net^{B,\epsilon/N}(|\phi_j\rangle) \leq -2 N \log\frac{\epsilon}{N}~,
\eest
thus the complexity of a separable state grows only at most linearily
with the number of qubits.
\end{es}

\begin{nota}
As we have seen, in general the growth of the complexity of a state
with the number of qubits is exponential, and this is substantially
different to what happens in the classical case where the upper bound
is linear with the number of bits. In this last example, though, we
see that the absence of entenglement re-establishes the classical
limit: entanglement thus proves again being a fundamental feature that 
distinguishes quantum objects from classical ones.
\end{nota}

\subsection{Complexity of graph states}

Graph states are multi-particle entangled states that can uniquely be
described by 
mathematical graphs, where the vertices of the graph take the role of
qubits and edges represent unitary operations between the relative
qubits \cite{marc}. 

Given a graph $G= (V,E)$ one cane easily prepare the corresponding
graph state using the following procedure:
\begin{enumerate}
\item[\emph{i.}] prepare all qubits in state $|+\rangle =
  (\snul+|1\rangle)/\sqrt{2}$;
\item[\emph{ii.}] when there is an edge between two vertices $k$ and
  $l$ apply a controlled-phase gate between the two qubits.This
  actually means applying to the two qubits a transformation of the
  form: $\displaystyle{U_{kl}= \mathrm{e}^{{-i\frac{\pi}{4}
      ({\mathbf{1}}^{(k)}-\sigma_z^{(k)})
      ({\mathbf{1}}^{(l)}-\sigma_z^{(l)})}}}$~.
\end{enumerate}
The resulting state $\spsi_G$ will be an entangled state uniquely
described by the graph $G$.

Once fixed the number $N$ of vertices, there are at most
$2^{N(N-1)/2}$ different graphs $G_1,G_2,\cdots,G_{2^{N(N-1)/2}}$ (each
vertex can in fact be connected 
or not to each other vertex by an edge). Correspondingly, given $N$
qubits, we can build at most $2^{N(N-1)/2}$ different graph states
$\spsi_{G_1}, \spsi_{G_2}, \cdots, \spsi_{G_{2^{N(N-1)/2}}}$. 

Once we define some sort of lexicographic order in the set of all
graphs (or equivalently in that of the graph states) only $\mco(N^2)$  
bits are sufficient to specify a determinate state. This value
constitutes an upper bound for the complexity of a graph state. 

We
will now show that this same value can be obtained for the algorithmic 
complexity of a graph state. As seen above in the more general case,
the upper bound for the complexity of a state is obtained finding a
bound for the length (and thus for the complexity) of the
characterizing string. In the case of graph states this bound is
obtained in a simple way: if $N$ is the number of vertices of the
graph, the maximum number of edges is $N(N-1)/2$. This implies that
a corresponding graph state can be obtained applying to the $N$ qubits
at most $N+N(N-1)/2$ gates ($N$ Hadamard gates, needed to initially
prepare all the qubits in state $|+\rangle$, and $N(N-1)/2$
controlled-phase gates). If $G$ has $N$ vertices, we have thus:
\beq
K_\net(\spsi_G) = K_\cl(\hat{\bomega}^B(\spsi_G)) \leq
l(\hat{\bomega}^B(\spsi_G)) \lesssim N+N(N-1)/2\lesssim N^2 .
\label{graphexact}
\eeq

\vspace{0.2cm}

As there is no dependence on the precision parameter,
Eq. (\ref{graphexact}) holds true only if we have the possibility to 
reproduce exactly the controlled-phase and the Hadamard gates. 

It is nevertheless possible to obtain a (more general) upper bound for
the complexity  of this family of states, valid also in the case in
which our basis is not of the above type. In order to do this we will
use some considerations regarding the complexity of
sentences, seen in section \ref{sentences}.

In order to prepare a graph state only two different types of gates,
 controlled-phase and Hadamard, are sufficient. If these gates cannot
 be reproduced exactly by those in our basis then, naturally, the
 wanted state can be prepared only with finite precision. As we have
 seen above, we need at most $\mco(N^2)$ of these gates to prepare any
 arbitrary graph state. To guarantee that the desired state is
 prepared with precision $\epsilon$ it is thus enough to reproduce
 each gate with  precision $\epsilon/N^2$. From the Solvay-Kitaev
 theorem we  know that implies we need to use $\mco(-\log
 \frac{\epsilon}{N^2})$ 
 gates to simulate each Hadamard (or controlled-phase) gate. Once we
 code the  circuit, to each Hadamard (or controlled-phase) will
 correspond the same $\mco(-\log\frac{\epsilon}{N^2})$-letter
 word. Using  Eq. (\ref{sentcompl}) (where $\#D=2$)\footnote{Actually
 the  cardinality of $D$ can be larger than 2, as there will be some
  words that correspond to operations such as ``new line'' or
 similar, but it will always be independent of $N$.} we obtain:
\beq
K_\net(\spsi_G)\leq N^2-\log\frac{\epsilon}{N^2}~.
\eeq

Different is the case of \emph{weighted graph states} \cite{lorenz}. These
states are generalizations of graph states, in which every edge is
specified by a 
(different) phase. The procedure to prepare these states is analogous
to that illustrated for graph states; the only difference is that, in
this case, whenever two vertices $k$ and $l$ are connected, one must
now apply a transformation of the form
$\displaystyle{U_{kl}= \mathrm{e}^{-i\frac{\phi_{kl}}{4}
      (\mathbf{1}^{(k)}-\sigma_z^{(k)})
      (\mathbf{1}^{(l)}-\sigma_z^{(l)})}}$~.

While the total number of gates in the circuit is still at most
$\mco(N^2)$, in this case it is not sufficient to consider only
Hadamard and controlled-phase gates as, in principle, each phase-gate
could be different. When preparing a weighted graph state it is thus
necessary to reproduce $\mco(N^2)$ different gates with precision
$\mco(- \log\frac{\epsilon}{N^2})$. Again we can obtain an upper bound
for the complexity of these states by using Eq. (\ref{sentcompl}),
only that this time the size of the dictionary does depend on $N$:
$\#D\sim N^2$. We have thus:
\beq
K_\net(\spsi_{\textit{W.G.}})\leq N^2
-N^2\log\frac{\epsilon}{N^2}\simeq-N^2\log\frac{\epsilon}{N^2}~.
\eeq

\section{Conclusions and outlook}

In this paper we have introduced a new definition for the algorithmic
complexity of quantum states. We have defined the complexity of a
quantum state as the description complexity of its experimental
preparation via a quantum circuit. We have investigated the relation
between the Shannon entropy of a source and the algorithmic complexity
of the emitted message.  We could also straightforwardly apply this
definition to find upper 
bounds for a number of interesting cases. We have seen a relation
between entanglement and algorithmic complexity: in particular we have
seen that the absence of entanglement reduces the upper bound for the
algorithmic complexity to the classical one.

While we have studied the algorithmic complexity of some classes of
states, one could pursue this investigation analysing
other states, for example states that appear in the context of quantum
phase transition and quantum adiabatic computation \cite{latorre}. Recent
results \cite{ham} suggest that it could be possible to construct
stricter upper bounds for preparation complexity of these
states. From a broader perspective, it would be interesting to
investigate further connections 
between algorithmic complexity and entanglement.

\vspace{0.6cm}

\noindent
\emph{We want to thank Fabio Benatti for useful discussions and
  comments. This work has been supported in part by the Deutsche
  Forschungsgemmeinschaft and the European Union
  (IST-2001-38877,-39227).} 

\section*{Appendix A: Connection with Vitanyi's Complexity}

In \cite{vitanyi} the author proposes a definition of quantum
algorithmic complexity based on quantum Turing machines. In
particular, the algorithmic complexity of a quantum state is given by
the following expression:
\beq
K_{\textrm{Vit}}(\sphi)=\min\{l(p)+\lceil -\log(|\langle\psi\sphi|^2)
\rceil \}~,
\eeq
where the minimum is taken over all programs $p$ (in classical bits)
running on a universal Turing machine $\mcu$ and such that
$\mcu(p,|\phi_0\rangle)=|\psi\rangle$ (where $|\phi_0\rangle$ is the
initial state of the computer), and $l(p)$ is the (classical)
length of the program. The complexity of $\sphi$ is therefore
constituted of two separate terms: the length of a program describing
an approximation $\spsi$ to $\sphi$ and a term penalizing for a bad
approximation.

The author shows that, if $\sphi\in \mcq_N$, then $2N$ is an upper
bound for $K_{\textrm{Vit}}(\sphi)$. This is easilly obtained
considering the projections of $\sphi$ over the vectors of a basis:
being $\sphi$ normalized, there exists at least one vector $|e_j\rangle$
such that $\vert \langle e_j\sphi\vert^2\geq1/2^{N}$: the length of a
program that 
gives $|e_j\rangle$ as output is at most $n$ and the corresponding
value for $l(p)+\lceil -\log(|\langle\psi\sphi|^2)\rceil$ is exactly
$2N$.  

The required upper bound for the complexity of a state $\sphi$ is thus 
obtained when the output state $\spsi$ satisfies the condition
$|\langle\psi\sphi|^2 \geq 1/2^N$. Such condition can be easily
rewritten in terms of a precision parameter $\epsilon_{\mathrm{Vit}}$ as:
$|\langle\psi\sphi|^2 \geq 1 - \epsilon_{\mathrm{Vit}}$ with
$\epsilon_{\mathrm{Vit}}  = 1 -1/2^N$. Rewriting the expression for
the upper bound in this particular case (and considering the case
$1/2^N \ll 1$) we have:
\best
K^{\epsilon_{\mathrm{Vit}}}(\sphi)\leq 2^N \log\epsilon_{\mathrm{Vit}}
  +N \simeq N~.
\eest
To find this expression it is necessary to consider the complete
expression for the upper bound given in Eq. (\ref{PrelCondLin}). We
see thus that the definition given by the author satisfies the
preliminary condition introduced in section
\ref{preliminary}. 

With this value of the precision parameter $\epsilon$, for the algorithmic 
complexity of a state we have:
\best
K^{\epsilon_{\mathrm{Vit}}}_\net(\sphi)\leq N^2 2^N
\log\epsilon_{\mathrm{Vit}} \simeq N^2~.
\eest
This bound is larger than the one that characterizes
Vitanyi's complexity; such a difference is only polynomial.

\section*{Appendix B: Quantum algorithmic complexity of a classical
  string} 

We naturally want to see if our definition is coherent with that of
classical Kolmogorov complexity. In order to do this we consider a
classical $N$-bit string $\boldx=x_{i_1} x_{i_2} \cdots x_{i_N}$
(where $x_{i_j} \in \{0,1\}$). 

In order to use a procedure built to characterize complexity of
quantum state it is necessary to ``translate'' the classical string
into a quantum state; this is easily done simply considering the state 
$|\boldx\rangle = |x_{i_1}\rangle |x_{i_2}\rangle \cdots
|x_{i_N}\rangle$, with $|x_{i_j}\rangle \in \{|0\rangle, |1\rangle\}$.

Considering the nature of this state it is trivial to see that it can
be obtained from the initial state $|0\rangle|0\rangle\cdots
|0\rangle$ by simply applying one-qubit \emph{NOT} gates to the bits
corresponding to $|x_{i_j}\rangle = |1\rangle$ and leaving invaried
(applying the identity transformation) the bits $|x_{i_j}\rangle =
|0\rangle$. 

To encode these particular circuits, thus, 3 symbols are
enough\footnote{In principle, in this case we need not define a
  complete gate basis as the only NOT gate is sufficient.}:
\begin{align*}
I &\leftrightarrow \mathrm{identity}\\
N &\leftrightarrow \mathrm{NOT~gate}\\
L &\leftrightarrow \mathrm{NEW LINE}~.
\end{align*}

\noindent
Let us consider now a simple example:
\bcen
\begin{tabular}{l c c c c c c c c c c}
Classical string & $\boldx$ & = & 1 & 0 & 1 & 1 & 0 & 1 & 0 & 0\\
Quantum state & $|\boldx\rangle$ & = & $|1\rangle$ & $|0\rangle$ &
$|1\rangle$ & $|1\rangle$ 
& $|0\rangle$ & $|1\rangle$ & $|0\rangle$ & $|0\rangle$.
\end{tabular}
\ecen

\bcen
\begin{figure}[h]
  \centering
  \includegraphics*[angle=0,
        scale=0.6]{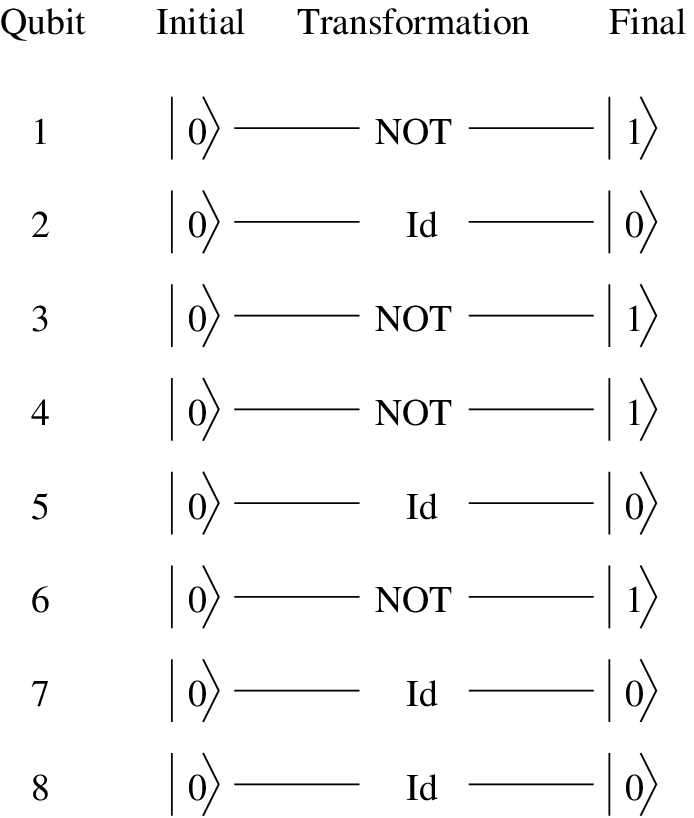} 
\end{figure}
\ecen

After coding, then, we have the following string representing the
circuit:
\bcen
$\bomega(\mcc) = N~L~~~I~L~~~N~L~~~N~L~~~I~L~~~N~L~~~I~L~~~I~L$
\ecen
(where, to help visualization, we have put larger spaces after line
breaks $L$). It is evident that there is a direct correspondence
between $\bomega(\mcc)$ and the original classical string $\boldx$: in
fact, a program that reproduces $\boldx$ reproduces also $\bomega(\mcc)$
(with the agreement that $0\leftrightarrow I$ and $1\leftrightarrow N$) 
(it will be enough to add a \emph{constant} part that tells the
program to insert a line-break after every simbol). Thus it follows
immediatly that the (classical) Kolmogorov complexity of
$\bomega(\mcc)$ coincides with that of $\boldx$. Considering the
definition of the network complexity we have immediatly:
\beq
K_\net(|\boldx\rangle) \simeq K_{\textrm{Cl}}(\boldx)~.
\eeq
Naturally the exact procedure followed in the example can be applied
to any $N$-bit string $\boldx$ so the result is true in general.


\begin{thebibliography}{99}

\bibitem{ham}  D. Aharonov, W. van Dam, J. Kempe, Z. Landau, S. Lloyd,
  O. Regev, \emph{Adiabatic Quantum Computation is Equivalent to
  Standard Quantum Computation}, arXiv: quant-ph/0405098 (2004).  

\bibitem{tutti} C. H. Bennett, P. G\'acs, M. Li, P. M. B. Vitanyi,
  W. H. Zurek, \emph{Thermodynamics of Computation and Information
  Distance}, Proc. 25th ACM Symp. Theory of Computation, ACM Press
  (1993). 


\bibitem{vandam} A. Berthiaume, W. van Dam, S. Laplante, \emph{Quantum
  Kolmogorov Complexity}, arXiv: quant-ph/005018 (2000).

\bibitem{chaitin1} G. J. Chaitin, \emph{Information, Randomness \&
    Incompleteness}, World Scientific (1987).


\bibitem{lorenz} W. D\"ur, L. Hartmann, M. Hein, H. J. Briegel,
  \emph{Entanglement in Spin Chains and Lattices with Long-Range
  Interactions}, arXiv:quant-ph/0407075 (2004). 


\bibitem{gacs} P. G\'acs, \emph{Quantum Algorithmic Entropy}, arXiv: 
  quant-ph/0011046 v2 (2001).

\bibitem{gruenwald} P. Gr\"unwald, P. Vitanyi, \emph{Shannon
    Information and Kolmogorov Complexity}, arXiv:cs.IT/0410002 v1
    (2004). 

\bibitem{marc} M. Hein, J. Eisert, H. J. Briegel, \emph{Multi-party
    Entanglement in Graph States}, arXiv:quant-ph/0307130v6 (2004).

\bibitem{holevo} A. S. Holevo, \emph{Bounds for the Quantity of
    Information Transmitted by a Quantum Communication Channel},
    Problems in Information Tranmission., {\bf{9}}, 177 (1973).

\bibitem{kitaev} A. Y. Kitaev, \emph{Quantum computations: algorithms
    and error correction}, Russ. Math. Surv. {\bf{52}}, 1191 (1997).

\bibitem{kolmogorov} A. N. Kolmogorov, \emph{Three Approaches to the
  Quantitative Definition of Information}, Probl. Inf. Transmission,
  {\bf{1}}, 1 (1965).

\bibitem{latorre} J. I. Latorre, R. Orus \emph{Adiabatic quantum
    computation and quantum phase transitions},  Phys. Rev. A
    {\bf{69}}, 062302 (2004).

\bibitem{li}M. Li, P. Vitanyi, \emph{An Introduction to Kolmogorov
    Complexity and Its Applications}, Springer (1997).

\bibitem{nielsen} M. A. Nielsen, I. L. Chuang, \emph{Quantum
    Computation and Quantum Information}, Cambridge University Press
    (2000).

\bibitem{schumacher} B. Schumacher, \emph{Quantum coding},
  Phys. Rev. A {\bf{51}}, 4 (1995).


\bibitem{vitanyi} P. Vitanyi, \emph{Three Approaches to the
    Quantitative Definition of Information in an Individual Pure
    Quantum State}, arXiv: quant-ph/9907035 (2000).



\end{thebibliography}
\end{document}